\RequirePackage{fixltx2e}

\documentclass[aps,prc,showpacs,floatfix,twocolumn,nofootinbib,amsfonts,amsmath,amssymb,superscriptaddress,altaffillsymbol]{revtex4}

\usepackage[dvipdfm,colorlinks=true,citecolor=blue,linkcolor=blue,filecolor=blue,urlcolor=blue]{hyperref}
\usepackage{fixltx2e}
\usepackage{graphicx}
\usepackage{float}
\usepackage{dblfloatfix}
\usepackage{dcolumn}
\usepackage{amsmath}

\newcommand{\etal}{\emph{et~al.}}
\begin{document}

\title{Structure of $^{13}$Be probed via secondary beam reactions}

\author{G.~Randisi}
\altaffiliation{Present address: Instituut voor Kern- en Stralingsfysica, KU Leuven, B-3001 Leuven, Belgium.}
\affiliation{LPC-ENSICAEN, IN2P3-CNRS et Universit\'{e} de Caen, 14050 Caen cedex, 
France}
\author{A.~Leprince}
\affiliation{LPC-ENSICAEN, IN2P3-CNRS et Universit\'{e} de Caen, 14050 Caen cedex, 
France}
\author{H.~Al~Falou}
\altaffiliation{Present address: Universit\'{e} de Technologie et de Sciences Appliqu\'{e}es Libano-Fran\c{c}aise,
Tripoli, Lebanon.}
\affiliation{LPC-ENSICAEN, IN2P3-CNRS et Universit\'{e} de Caen, 14050 Caen cedex, 
France}
\author{N.A.~Orr}
\affiliation{LPC-ENSICAEN, IN2P3-CNRS et Universit\'{e} de Caen, 14050 Caen cedex, 
France}
\author{F.M.~Marqu\'es}
\affiliation{LPC-ENSICAEN, IN2P3-CNRS et Universit\'{e} de Caen, 14050 Caen cedex, 
France}
\author{N.~L.~Achouri}
\affiliation{LPC-ENSICAEN, IN2P3-CNRS et Universit\'{e} de Caen, 14050 Caen cedex, 
France}
\author{J.-C.~Ang\'{e}lique}
\affiliation{LPC-ENSICAEN, IN2P3-CNRS et Universit\'{e} de Caen, 14050 Caen cedex, 
France}
\author{N.~Ashwood}
\affiliation{School of Physics and Astronomy, University of Birmingham, Birmingham B15 2TT, UK}
\author{B.~Bastin}
\altaffiliation{Present address: GANIL, Caen, France.}
\affiliation{LPC-ENSICAEN, IN2P3-CNRS et Universit\'{e} de Caen, 14050 Caen cedex, 
France}
\author{T.~Bloxham}
\affiliation{School of Physics and Astronomy, University of Birmingham, Birmingham B15 2TT, UK}
\author{B.A.~Brown}
\affiliation{NSCL and Department of Physics and Astronomy, Michigan State University, East Lansing, MI~48824, USA}
\author{W.~N.~Catford}
\affiliation{Department of Physics, University of Surrey, Guildford GU2 7XH, UK}
\author{N.~Curtis}
\affiliation{School of Physics and Astronomy, University of Birmingham, Birmingham B15 2TT, UK}
\author{F.~Delaunay}
\affiliation{LPC-ENSICAEN, IN2P3-CNRS et Universit\'{e} de Caen, 14050 Caen cedex, 
France}
\author{M.~Freer}
\affiliation{School of Physics and Astronomy, University of Birmingham, Birmingham B15 2TT, UK}
\author{E.~de G\'{o}es Brennand}
\affiliation{PNTPM, CP-229, Universit\'{e} Libre de Bruxelles, B-1050 Brussels, Belgium}
\author{P.~Haigh}
\affiliation{School of Physics and Astronomy, University of Birmingham, Birmingham B15 2TT, UK}
\author{F.~Hanappe}
\affiliation{PNTPM, CP-229, Universit\'{e} Libre de Bruxelles, B-1050 Brussels, Belgium}
\author{C.~Harlin}
\affiliation{Department of Physics, University of Surrey, Guildford GU2 7XH, UK}
\author{B.~Laurent}
\altaffiliation{Present address: CEA, DAM, DIF, F-91297 Arpajon, France.}
\affiliation{LPC-ENSICAEN, IN2P3-CNRS et Universit\'{e} de Caen, 14050 Caen cedex, 
France}
\author{J.-L.~Lecouey}
\affiliation{LPC-ENSICAEN, IN2P3-CNRS et Universit\'{e} de Caen, 14050 Caen cedex, 
France}
\author{A.~Ninane}
\affiliation{Slashdev Integrated Systems Solutions, B-5030 Gembloux, Belgium}
\author{N.~Patterson}
\affiliation{Department of Physics, University of Surrey, Guildford GU2 7XH, UK}
\author{D.~Price}
\affiliation{School of Physics and Astronomy, University of Birmingham, Birmingham B15 2TT, UK}
\author{L.~Stuttg\'{e}}
\affiliation{IPHC, Universit\'{e} Louis Pasteur et IN2P3-CNRS, 67037 Strasbourg Cedex 02, France}
\author{J.~S.~Thomas}
\altaffiliation{Present address: Schuster Laboratory, The University of Manchester, UK.}
\affiliation{Department of Physics, University of Surrey, Guildford GU2 7XH, UK}

\date{\today}

\begin{abstract}

The low-lying level structure of the unbound neutron-rich nucleus $^{13}$Be has been investigated 
via breakup on a carbon target of secondary beams of $^{14,15}$B at 35~MeV/nucleon.  
The coincident detection of the beam velocity $^{12}$Be fragments and
neutrons permitted the invariant mass of the $^{12}$Be+$n$ and $^{12}$Be+$n$+$n$ systems
to be reconstructed.  
In the case of the breakup of $^{15}$B, a very narrow structure at threshold was observed
in the $^{12}$Be+$n$ channel. 
Analysis of the $^{12}$Be+$n$+$n$ events demonstrated that 
this resulted from the sequential-decay of the unbound $^{14}$Be(2$^+$) state
rather than a strongly interacting $s$-wave virtual state in $^{13}$Be, as had been 
surmised in stable beam fragmentation studies. 
Single-proton removal from $^{14}$B
was found to populate a broad low-lying structure some 0.7~MeV above
the neutron-decay threshold in addition to a less prominent feature at 
around 2.4~MeV.
Based on the selectivity of the reaction and a comparison with (0-3)$\hbar\omega$ shell-model calculations,
the low-lying structure is concluded to arise from closely spaced J$^\pi$=1/2$^+$ and 5/2$^+$ resonances (E$_r$=0.40$\pm$0.03 
and 0.85$^{+0.15}_{-0.11}$~MeV), whilst the broad higher-lying feature is a second 5/2$^+$ level (E$_r$=2.35$\pm$0.14~MeV). 
Taken in conjunction with earlier studies, it would appear that the lowest 1/2$^+$ and 1/2$^-$ levels
lie relatively close together below 1~MeV.

\end{abstract}


\maketitle

\section{Introduction}
\label{sec:Introduction}

The light nuclei have long provided a test bench for our understanding of nuclear structure.  Indeed, one 
of the central themes of present day studies --- the evolution of shell-structure with 
isospin --- may arguably be 
traced back to the seminal work of Talmi and Unna on the ordering of the $\nu 2s_{1/2}$--$\nu 1p_{1/2}$
levels in $^{11}$Be \cite{Tal60}.
With the advent of radioactive beams it has become possible to explore experimentally isotopic chains 
in the light mass region to the very limits of stability and beyond, thus providing for more stringent tests of
nuclear structure models.  

In this context the beryllium isotopes are of particular interest as they exhibit a rich variety of structural phenomena, ranging from the $\alpha$-clustering
of $^{8}$Be and the deformed 2$\alpha$-X$n$ ``molecular'' character of $^{9,10}$Be 
\cite{Oer97,Fre06}, through the 
single-neutron halo of $^{11}$Be \cite{Mil83} and the breakdown of the N=8 $p$--shell closure of $^{12}$Be \cite{Bar76,Nav00,Pai06} to the Borromean two-neutron halo of $^{14}$Be \cite{14Behalo,14Behalo1,Suz99}.
The N=9 member of the Be isotopic chain, $^{13}$Be, which is neutron unbound \cite{refunbound13Be,refunbound13Be1}, 
thus presents an intriguing study.  First, in terms of the halo of $^{14}$Be, the level structure of $^{13}$Be is critical in establishing the  
$^{12}$Be+$n$ interaction, an essential component in three-body modelling \cite{Des95,Tho96}.
Second, from a structural point of view, the dominant $\nu 2s_{1/2}$ ground-state
configurations of the less exotic N=9 isotones $^{14}$B \cite{Gui00,Sau00,Sau04,Bed13} and $^{15}$C \cite{Sau00,Sau04,Gos75} suggest 
that the lowering of the 
$\nu 2s_{1/2}$ single-particle orbital with respect to the $\nu 1d_{5/2}$ responsible 
for this \cite{Tal60,Mil01} may well also 
generate an inversion of the 1/2$^+$ and 5/2$^+$ levels in $^{13}$Be.  Such behaviour, whilst in line with the
dominant $2s_{1/2}^2$ valence neutron structure of $^{14}$Be \cite{Lab01,Sim07,Ili12}, is
complicated by the deformed \cite{Iwa00} mixed (0+2)$\hbar\omega$ character of 
$^{12}$Be \cite{Bar76,Nav00,Pai06} which, as posited by some models \cite{Lab99,Bla10,Eny12}, would suggest that there is also a low-lying 1/2$^-$ state
in $^{13}$Be.  In the 
work presented here, we concentrate on locating the 
$\nu 2s_{1/2}$ and $\nu 1d_{5/2}$ levels and determining the character of the former.

Beginning with the first indication of an unbound resonance, lying around 1.8~MeV above the
$^{12}$Be+$n$ threshold (width $\Gamma$=0.9$\pm$0.5~MeV) \cite{Ale83}, $^{13}$Be has been the object of a
range of measurements employing both stable \cite{Ale83,Ost92,Bel98,Tho00,Chr08} and radioactive beams
\cite{Kor95,Sim07,Kon10,Mar01,Lec04,Aks13}, including pion absorption \cite{Gor98}.  
The heavy-ion multinucleon transfer studies all agree on
the existence of a well defined resonance around 2~MeV above the $^{12}$Be+$n$ threshold
\cite{Ale83,Ost92,Bel98}, the intrinsic width ($\Gamma \approx$ 0.3~MeV) of which was compatible with an $\ell$=1 or 2 assignment \cite{Ost92}.  Strong arguments for a 5/2$^+$ assignment were made based on the comparison of the 
($^{11}$B,$^{12}$N) reaction on $^{12}$C and $^{14}$C, leading to $^{11}$Be and $^{13}$Be, whereby 
the well-known $^{11}$Be 1.78~MeV 5/2$^+_1$ state 
was the most strongly populated in the former case \cite{Bel98}.  A weakly populated peak was also 
observed at 0.80$\pm$0.09~MeV above threshold in $^{14}$C($^{11}$B,$^{12}$N), which, based again on comparison with the same reaction on $^{12}$C, was tentatively assigned J=1/2.  It may be noted, however, that such heavy-ion multi-nucleon transfer reactions with very negative Q-values do not favor low transferred angular momentum \cite{Bri72,Cat89}.  Indeed, as is well evidenced by
attempts \cite{Boh93,You94,Cag99} to locate the now well-known threshold $\nu 2s_{1/2}$ state in $^{10}$Li, it is extremely
difficult to populate such states in this manner.  In addition, given that stopped pion-absorption 
reactions do not appear to readily populate 
$\ell$=0 states \cite{Pion-noell-zero} and the preliminary report of a resonance 0.65$\pm$0.10~MeV
above threshold in the $^{14}$C($\pi^-$,p) reaction \cite{Gor98}, it may be concluded 
that there is quite probably a 1/2$^-$ level around 0.8~MeV above threshold.  It is worthwhile noting 
that, as opposed to the $^{12}$Be+$n$ invariant mass studies, which can be influenced by bound excited
states of the $^{12}$Be fragment, missing mass measurements provide unambiguous
determinations of the $^{13}$Be energies.  

Two other stable beam studies have been undertaken in which the relative velocity spectra for
$^{12}$Be fragments and neutrons following fragmentation of beams of $^{18}$O \cite{Tho00} and $^{48}$Ca \cite{Chr08} were investigated.  In both cases very narrow structures centred at zero were observed and
interpreted as arising from a strongly interacting virtual $s$-wave state (scattering length $a_s<-$10~fm).    

Turning to the radioactive beam reaction studies, both proton and neutron removal (or ``knockout'') 
from $^{14}$B \cite{Lec04} and $^{14}$Be \cite{Sim07,Kon10,Aks13} respectively have been investigated. 
The proton removal study of Lecouey \cite{Lec04}, which reports our first efforts to study the C($^{14}$B,$^{12}$Be+$n$) reaction, observed in the decay energy spectrum a very broad structure ($\Gamma \approx$ 1.3~MeV) centred 
some 0.7~MeV above threshold together with a narrower peak ($\Gamma \approx$ 0.4~MeV)
at around 2.4~MeV which were identified, based on the selectivity of the reaction,
as 1/2$^+$ and 5/2$^+$ resonances.  The subsequent neutron-knockout studies \cite{Sim07,Kon10,Aks13} 
exhibited $^{12}$Be+$n$ invariant
mass (or decay energy) spectra with a somewhat narrower low-lying structure at $\sim$0.5~MeV and a
broader feature at around 2--3~MeV.  At intermediate energies with a proton target  
and guided by the reconstructed 
$^{12}$Be+$n$ transverse momentum distributions, a $p$-wave resonance at 0.51$\pm$0.10~MeV 
($\Gamma$=0.45$\pm$0.3~MeV) and a very broad ($\Gamma$=2.4$\pm$0.2~MeV) $d$-wave structure were
identified \cite{Kon10}.  These features were superposed on an underlying 
distribution which was modelled as a very weakly
interacting $s$-wave virtual state (scattering length a$_s\approx$~$-$3~fm).  At high 
beam energy using a carbon target
the interpretation was also guided by the $^{12}$Be+$n$ transverse momentum distributions
as well as $^{12}$Be--$n$ angular correlations \cite{Sim07}.  In this case, a more complex interpretation of the 
invariant mass spectrum was proposed in which $^{13}$Be states at 2.0 (assigned J$^{\pi}$=5/2$^+$) and 3.0 MeV (1/2$^-$) have neutron decay branches to $^{12}$Be$^*$(2$^{+}_{1}$,$0^+_2$,$1^-_1$).  In addition, a significant contribution from a very weakly interacting $s$-wave virtual state (a$_s\approx$~$-$3~fm) was postulated. 
Very recently, improved data recorded at very high beam energies using a proton target have been
reported \cite{Aks13} and a detailed analysis, taking into account earlier studies (in particular Refs~\cite{Lec04,Kon10}), suggests that the decay energy spectrum is dominated by resonant $s$-wave strength,
especially at low decay energy, in addition to a 5/2$^+$ state at 2.0~MeV and other possible weaker higher-lying states at around 3 and 5~MeV \cite{Aks13a}.
Finally, despite a relatively high background, 
an early investigation of the d($^{12}$Be,p) transfer reaction also indicated the presence of a state at 
around 2~MeV \cite{Kor95}.

From a theoretical point of view, approaches ranging from the shell model  
\cite{Kon10,Pop85,For13} to various few-body 
calculations \cite{Des95,Tho96,Lab99,Bla10,Des94,Pac02,Tar04}, including 
deformation \cite{Ham08}, together with the 
Relativistic Mean Field \cite{Ren97} and Antisymmetrised Molecular Dynamics \cite{Eny12} models  
have been employed to explore the structure of $^{13}$Be.  Apart from general agreement on the existence of a
$\nu 1d_{5/2}$ resonance well removed from the $^{12}$Be+$n$ threshold, no systematic conclusions may be drawn regarding any
lower-lying levels.  It is interesting to note that the 5/2$^+$ level, as observed in the 
$^{13}$C($^{14}$C,$^{14}$O)$^{13}$Be reaction at 2.01$\pm$0.05~MeV, has served as a reference point, together with the
constraint provided by the $^{14}$Be binding energy, for the majority of the few-body 
calculations \cite{Des94,Des95,Lab99,Pac02}.

\begin{figure}[t!]
 \begin{center}
  \includegraphics[width=\columnwidth]{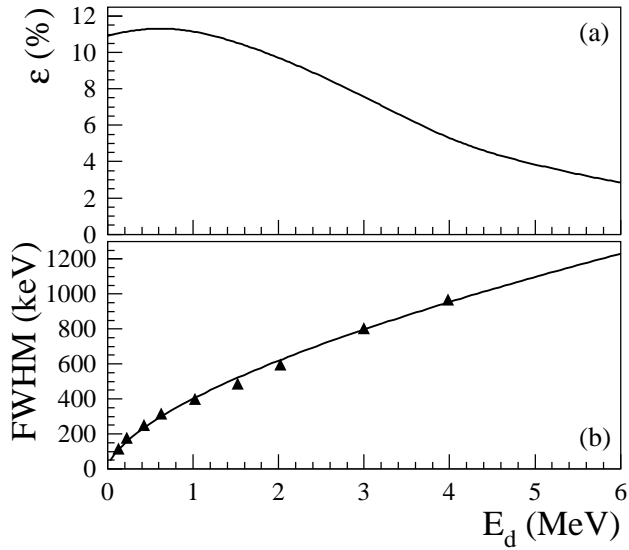}
  \end{center}
  \caption{\small \rm (a) Simulated efficiency for the detection of $^{12}$Be+$n$ pairs as a function of
  decay energy (E$_d$). (b) Simulated resolution of the reconstructed $^{12}$Be+$n$ decay energy.}
   \label{fig:Response}
\end{figure}

In the present paper we report on the investigation of $^{13}$Be using single-proton removal from $^{14}$B
and breakup of $^{15}$B at intermediate energies (35~MeV/nucleon).  As argued elsewhere \cite{Zin95,Che01,Lec09,AlF11}, proton removal  presents the advantage of preferentially populating states whereby the neutron configuration remains little perturbed
with respect to that of the projectile.  As such, it is expected that by employing a
$^{14}$B beam here final states in $^{13}$Be with $\nu 2s_{1/2}$ and $\nu 1d_{5/2}$
configurations should be preferentially populated \cite{Gui00,Sau00,Sau04,Bed13}.  In principle, 
breakup of $^{15}$B, which involves proton and neutron removal,
has the potential to populate a broader variety of states.  However, as will be seen, the $^{12}$Be+$n$
channel is dominated by events of very low relative energy resulting from the in-flight decay of the unbound 2$^+_1$
state of $^{14}$Be, which were interpreted in earlier stable beam fragmentation experiments as
arising from a strongly interacting threshold $s$-wave virtual state in $^{13}$Be \cite{Tho00,Chr08}.  

\section{Experiment}

The $^{14,15}$B secondary beams with mean energies of 35~MeV/nucleon were
produced via the reaction on a 3.1~mm thick Be target of a 55~MeV/nucleon $^{18}$O primary beam supplied
by the {\sc ganil} coupled-cyclotron facility.
The beam velocity reaction products were analysed and purified 
using the {\sc lise}3 fragment separator employing an achromatic degrader \cite{Ann87}.
Two separate settings of the separator were used for each of the beams so as to optimise the intensities
($^{14}$B, $\sim$1.3$\times$10$^4$~pps and $^{15}$B, $\sim$8$\times$10$^3$~pps) and purities ($>$95\%).
  
\begin{figure}[t!]
 \begin{center}
  \includegraphics[width=\columnwidth]{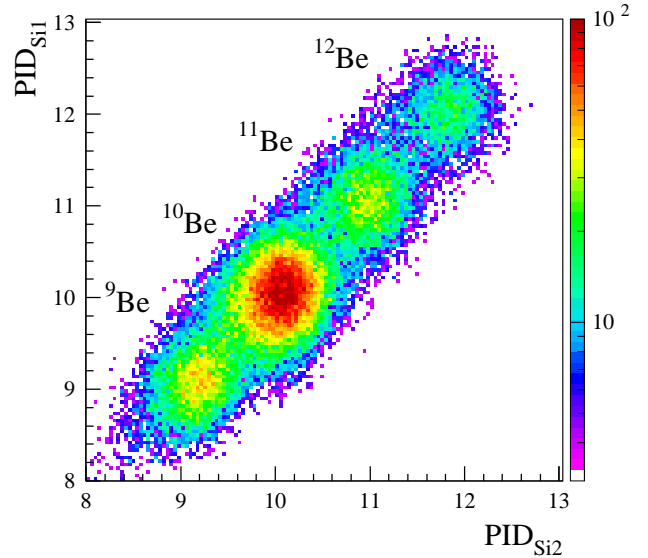}
  \end{center}
  \caption{\small \rm (Color online) Particle identification (PID) derived from the Si-Si-CsI detector telescope for the beryllium isotopes in coincidence with beam velocity neutrons for the C($^{14}$B,$^A$Be+$n$) 
  reaction (see text).}
   \label{fig:PID}
\end{figure}

A time-of-flight measurement performed using the signal derived from the cyclotron RF and a 
thin (100~$\mu$m thick) plastic scintillator detector mounted just upstream of the first beam tracking detector (see below) allowed the $^{14,15}$B ions to be 
easily separated in the off-line analysis from the remaining 
contaminants.  The energy spread of the $^{14,15}$B beams, as defined by the settings of 
the {\sc lise} spectrometer, was 1.5\% ($\Delta E/E$).  The effect on
the relative energy between the $^{12}$Be  fragment and neutron from the in-flight decay of 
$^{13}$Be was negligible compared to the overall resolution and no event-by-event
correction based on the measured beam particle velocities was required. 

The secondary beam was delivered onto a 190~mg/cm$^2$ $^{nat}$C target.
Owing to the limited optical qualities of the secondary beams (beam spot size $\sim$10~mm diameter),
two position-sensitive drift chambers \cite{Mac98} located just upstream of the 
secondary reaction target and separated by 55~cm were
used for tracking.  The fast timing signal provided by the thin scintillator detector was used
as the timing reference for the drift chamber position measurements.  The impact point of the beam 
on target was subsequently reconstructed event-by-event with a resolution of $\sim$1.5~mm ({\sc fwhm}).

The charged fragments from the reactions were detected and identified
using a Si-Si-CsI(Tl) telescope centred at 0$^\circ$
downstream of the secondary target. 
The two 500~$\mu$m thick (50$\times$50~mm$^2$) silicon detectors, each comprising 16 resistive strips
(3~mm wide), were mounted
such that the strips of the first detector provided for a measurement of position in the vertical 
direction whilst those of the second detector the horizontal position.  These detectors were located
15~cm downstream of the target and the impact point along each
strip was determined with a resolution of 1~mm ({\sc fwhm}), significantly better than the individual strip width.
In addition to the energy-loss measurements furnished by the silicon detectors, the residual energy
of each fragment was determined from the signals derived from an array of 16 25$\times$25$\times$25~mm$^3$ CsI(Tl) crystals which were located 30~cm downstream of the target so as to subtend the same solid angle
as the Si detectors.  The measurement of the total energy deposited in the telescope by the charged fragments
was calibrated using
a mixed secondary beam, which included $^{12}$Be, and for which the energy spread was limited to 
0.1\%.  A series of measurements was made over a range of rigidity settings 
of the {\sc lise} spectrometer such that
the $^{12}$Be calibration points covered the range of energies expected from the in-flight decay 
of $^{13}$Be. 
The total energy resolution
of the telescope was determined to be some 2\% ({\sc fwhm}) \cite{Ran11,Lep09}.
In addition to the measurements of breakup on the C target, data were also acquired
with the target removed so as to ascertain the contribution arising from interactions in the
telescope.  As the reactions of interest involve charge changing with respect to the projectile, this contribution was found to be negligible, as expected \cite{Mar96}.

\begin{figure}[t!]
 \begin{center}
  \includegraphics[width=\columnwidth]{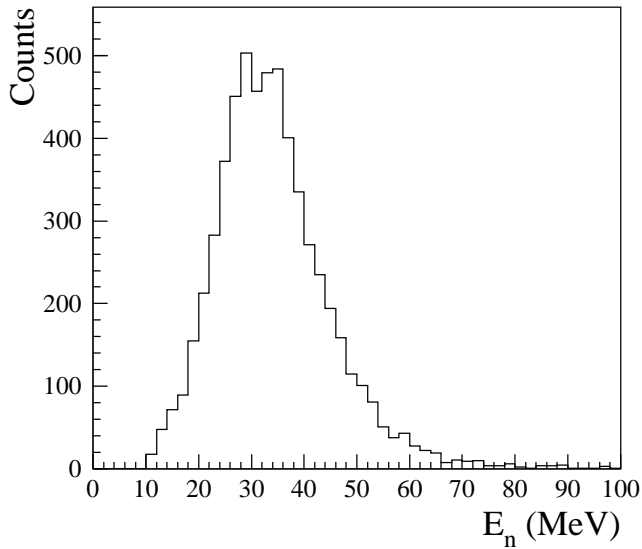}
  \end{center}
  \caption{\small \rm Neutron energy spectrum for events with $E_n>$11~MeV detected for 
  the C($^{14}$B,$^{12}$Be+$n$) reaction.}
   \label{fig:En}
\end{figure}  

\begin{figure}[t!]
 \begin{center}
  \includegraphics[width=\columnwidth]{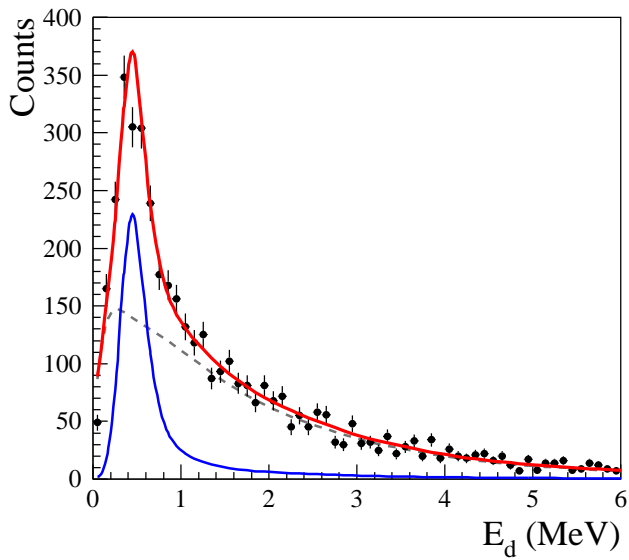}
  \end{center}
  \caption{\small \rm (Color online) Reconstructed $^6$He+$n$ decay energy for the C($^{14}$B,$^6$He+$n$) reaction.
  The results of a fit using the simulated lineshape for the known $^7$He 3/2$^-$ ground-state resonance (blue) together with an uncorrelated distribution generated by event 
  mixing (dashed line, see text) is shown.}
   \label{fig:7He}
\end{figure}

The neutrons were detected using 90 elements of the {\sc demon} liquid scintillator detector array \cite{Mos94,Til95}.  The use of such an array enabled the ambient $\gamma$-ray background to be eliminated using pulse shape discrimination (prompt reaction induced $\gamma$-rays can also be rejected using the time of flight), allowing, importantly, a very low threshold on the deposited energy (500~keVee) to be 
imposed, thus maximising the efficiency, rather than the $\sim$6~MeVee required to otherwise veto such events.  The intrinsic detection efficiency of each module was $\sim$35\% at 35 MeV \cite{Lab99thesis}. The overall
time-of-flight resolution, including the contribution of the thin plastic start detector, 
was 1.5~nsec ({\sc fwhm}).

As in previous experiments \cite{Pai06,Lab01,Mar01,Lec04,Mar00b}, the array was
arranged in a staggered multi-wall type (or ``zigzag'') configuration \cite{Mar00}
covering polar angles up to 27$^\circ$ in the laboratory frame \cite{Lec02,Ran11}\footnote{The single-wall configuration presented in the review of Ref. \cite{Bau12} was only utilised in the very first radioactive-beam experiment employing {\sc demon} \cite{Mar96}.}.  This arrangement
provided for a good detection efficiency for the $^{12}$Be+n reaction channel out 
to $\sim$5~MeV relative energy (Figure~\ref{fig:Response}) and good granularity --- the angular resolution 
in the neutron detection\footnote{Ranging from 1.4--2.6$^{\circ}$ depending on the distance of the module from the target.}
being the major contribution to the final reconstructed $^{12}$Be+n decay energy resolution (Figure~\ref{fig:Response}) \cite{Lec02}.
This configuration was also optimised to provide, in the case of multi-neutron detection (here $^{12}$Be+n+n), for the
rejection of cross talk (events arising from the scattering of a neutron from 
one detector into another with the deposited energy registered above 
threshold in both) in the off-line analysis using kinematic considerations coupled with
the energy deposited in the modules \cite{Mar00}.  
Finally, it may be noted that the rate of events for which
a neutron was first scattered without detection by a module (or non-active element in the setup) and then detected in another module was estimated, via simulations, to be considerably less than 5\% of the
total number of $^{12}$Be+n events and produced negligible distortions in the reconstructed decay 
energy spectra, as further confirmed by the measurement of the $^7$He ground state (see below).

The neutron velocity was derived from the time
of flight measured between the thin plastic scintillator detector and 
{\sc demon} module(s) which registered an event.  The neutron energy
spectra were dominated by beam velocity neutrons produced in the projectile breakup --- the relatively weak low-energy component arising from neutrons evaporated from the 
target was removed in the off-line analysis by imposing a low-energy threshold of 
11~MeV \cite{Ran11,Lep09} (Figure~\ref{fig:En}).  

\section{Results}

The identification of the beryllium fragments in coincidence with fast neutrons ($E_n>$11~MeV) 
is shown in Figure~\ref{fig:PID} where two particle-identification parameters (PID) were constituted 
using the energy-losses from each of the two silicon detectors together with the residual energy derived 
from the CsI(Tl) crystals \cite{Lec02}.  As may be seen, the $^{12}$Be fragments are clearly separated from the more prolific lighter mass isotopes.   Furthermore, as expected, the neutrons detected in coincidence with the $^{12}$Be fragments exhibit energies centred around 35~MeV (Figure~\ref{fig:En}).

The decay (or relative) energy, $E_d$ was derived from the measured four-momenta of the $^{12}$Be 
fragment and neutron, 

\begin{equation}\label{equa:decayen}
E_{d}  =  M_{inv} - (m_{f} + m_{n})
\end{equation} 

\noindent where $m_{f}$ and $m_{n}$ are the masses of the charged fragment and the neutron respectively,
and the invariant mass, $M_{inv}$ is given by,

\begin{equation}\label{equa:invariant}
M_{inv}^{2} = (\epsilon_{f} + \epsilon_{n})^{2}-
(\vec{p}_{f}+\vec{p}_{n})^{2} 
\end{equation}  
 
\noindent where $\epsilon_{f}$ and $\epsilon_{n}$ are the total (kinetic and rest mass) energies of the fragment and neutron.

Given the complex nature of the setup employed here, careful attention
was paid, as in previous experiments of this type which we have carried out \cite{Lab01,Pai06,Lec09,AlF11,AlF07}, to understand its effects, in 
particular on the reconstructed decay energy.
To do so a detailed simulation which took into account
the characteristics (resolution, intrinsic efficiency and geometry) of all the detectors as well as the beam characteristics
and the effects of the target and the reaction was developed.    
Owing to the granular character of the neutron-detection array, an important element in  
correctly describing the response of the setup is, as noted in our earlier work \cite{Lec09,Lec02}, the 
transverse momentum distribution of the decaying neutron unbound system --- here $^{13}$Be.
The intrinsic widths of the $^{13}$Be transverse 
momentum distributions for both the C($^{14}$B,$^{12}$Be+n) and C($^{15}$B,$^{12}$Be+n) reactions were 
thus adjusted within the simulations such that they matched 
the experimental widths (300 and 340~MeV/c {\sc fwhm} respectively) reconstructed from the
measured momenta of the $^{12}$Be fragment and neutron \cite{Ran11,Lep09}.
Importantly, the simulations were bench marked not only against codes developed independently as
part of earlier work \cite{Lec02,AlF07} but also through the ability to reproduce the well known 
$^{7}$He 3/2$^-$ ground-state resonance \cite{Til02} which was populated here in the C($^{14,15}$B,$^{6}$He+n) reactions 
\cite{Ran11,Lep09} (Figure~\ref{fig:7He}).

The results of the simulations for the response of the setup for the C($^{14}$B,$^{12}$Be+$n$) 
reaction are displayed in Figure~\ref{fig:Response} as a function of decay energy.  Very similar behaviour was found for 
the C($^{15}$B,$^{12}$Be+X$n$) reactions, except for the overall detection efficiency ($\varepsilon$)
which is lower by a factor of 10 for the two-neutron channel \cite{Lep09}.  The predicted efficiency for detecting 
a $^{12}$Be+$n$ pair is shown in the upper panel.  Importantly, the response is a smooth function
of decay energy exhibiting no features which could mimic a resonance-like state.  The gradual
fall off
from a maximum of some 12\% at around 1~MeV is in line with simple geometrical considerations based
on the angular coverage of the neutron array (the efficiency for the detection of the $^{12}$Be fragments
is essentially 100\%). 
The resolution in the reconstructed decay energy, which is dominated by the 
finite angular size of the individual {\sc demon} modules, was determined to 
vary as 0.40$\sqrt{E_{d}}$~MeV (lower panel, Figure~\ref{fig:Response}), with, for example, a resolution of close to
330~keV expected at 0.7~MeV.  It may be noted that owing to the slow falloff in the detection efficiency with
decay energy, the apparent position of peaks are slightly down shifted in energy for E$_d$ above 1~MeV \cite{Ran11}.

As described in our earlier work, in addition to the events arising from fragment+neutron
final-state interactions associated with the decay of the unbound states, care must be taken to
account for uncorrelated events \cite{Lec09,Lec02,AlF11,AlF07}.  Such events may arise from a number of sources.
In reactions involving proton only removal, such as C($^{17}$C,$^{15}$B+n) \cite{Lec09,Lec02} or, here, C($^{14}$B,$^{12}$Be+n), the non-resonant 
continuum may be populated via the fragment recoil effect \cite{Che01}
and scattering of the weakly bound projectile valence neutron by the target\footnote{In practice, 
given the limited resolution and decreasing detection efficiency at high decay energies, events 
corresponding to
very broad, weakly populated levels may also contribute.}.  In the case of breakup, whereby the
outgoing channel of interest has fewer neutrons than the projectile --- here C($^{15}$B,$^{12}$Be+n) and 
C($^{14}$B,$^{6}$He+n) --- an
additional contribution arising from the detection of the neutron(s) that do not correspond to the 
population of fragment+neutron states will also be present.  As discussed in detail and
demonstrated in Refs 
\cite{Lec09,Lec02}, the distribution of uncorrelated events may be 
estimated by event mixing using the measured fragment+neutron pairs, provided that care 
is taken to eliminate the effects of any ``residual'' correlations arising from the
resonant structures themselves \cite{Mar00b}. Importantly, this technique involves
no ad hoc assumptions or parameterisations and the
distribution so obtained incorporates explicitly the effects of the experimental response.  
As such it may be compared directly with the measured distribution in order to identify features arising from 
final-state interactions associated with the decay of unbound states \cite{Lec02}.  

Assuming that the form of the event-mixed distribution provides a good description of the uncorrelated events 
(and in practice this appears to be the case \cite{Lec09,Lec02,AlF11,AlF07}), the overall contribution
to the decay-energy spectrum must be determined. In the following, as in our earlier studies,
the contribution of the uncorrelated distribution to the invariant mass spectrum was allowed to vary 
within the fitting procedures used to derive the final results and associated uncertainties.
In addition, based on these fits, the ability to reproduce other observables, such as the $^{12}$Be+n
momentum distributions and the neutron energy spectrum, was verified \cite{Lep09,Ran11}.

\subsection{The C($^{14}$B,$^{12}$Be+$n$) reaction}

\begin{figure}[t!]
 \begin{center}
  \includegraphics[width=\columnwidth]{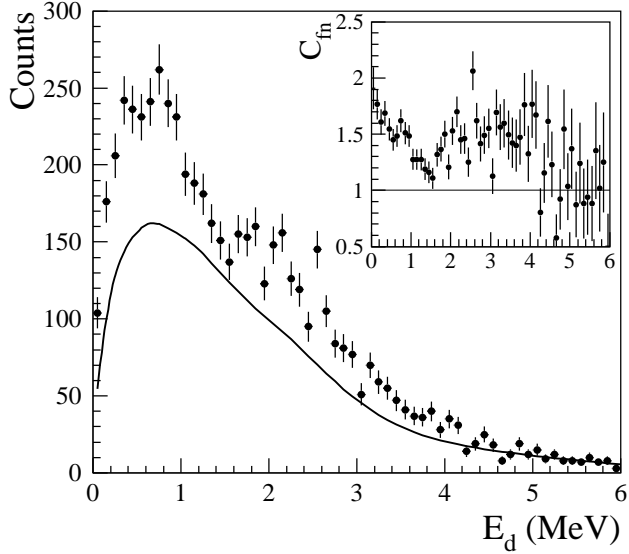}
  \end{center}
  \caption{\small \rm Reconstructed $^{12}$Be+$n$ decay energy for the C($^{14}$B,$^{12}$Be+$n$) reaction
  compared to the uncorrelated distribution (solid line, see text) normalised to the region above 4~MeV.  The inset
  shows the $^{12}$Be+$n$ correlation function, C$_{fn}$ (see text).}
   \label{fig:14B13BeEd}
\end{figure}

The reconstructed decay energy spectrum obtained for the C($^{14}$B,$^{12}$Be+$n$) reaction 
is shown in Figure~\ref{fig:14B13BeEd}.  For comparison, the uncorrelated distribution obtained via event
mixing and normalised to the region above 4~MeV, is included.  It is clear
that structures corresponding to $^{12}$Be+$n$ final-state interactions are present at
$\sim$1~MeV and below, as well as in the region around 2--3~MeV.  This is even more apparent in the inset of 
Figure~\ref{fig:14B13BeEd}, whereby 
the $^{12}$Be+$n$ correlation function, C$_{fn}$, 
obtained as the ratio of the measured and uncorrelated distributions, is displayed.

Before attempting to interpret the $^{12}$Be+$n$ decay energy spectrum, the
issue of bound excited states of $^{12}$Be should be addressed.  At present, $^{12}$Be is known to have 
three bound excited states -- the well established 2$^+_1$ state at 2.11~MeV \cite{Alb78}, together with
a 1$^-$ state at 2.7~MeV \cite{Iwa00} and an isomeric 0$^+_2$ level ($\tau_{1/2}\approx$300~nsec)
at 2.25~MeV \cite{Shi03,Shi07}.  The population of any of these levels will, as alluded to in the 
Introduction, provide for ambiguities in the interpretation of the $^{12}$Be+$n$ coincidences.  For example,
if a level in $^{13}$Be lying 2.30~MeV above the $^{12}$Be+$n$ decay threshold were to have
an appreciable branching ratio for decay to $^{12}$Be(2$^+_1$)+$n$, then the $^{12}$Be+$n$ decay
energy spectrum would exhibit, in addition to a peak at 2.30~MeV, another near threshold (E$_d$=0.19~MeV).
In the present experiment, no dedicated gamma detection was available.  It was possible, however, to 
obtain an estimate of the percentage of $^{12}$Be fragments in bound excited states (other than the 
isomeric state) using the {\sc demon} array.  

\begin{figure}[t!]
 \begin{center}
  \includegraphics[width=\columnwidth]{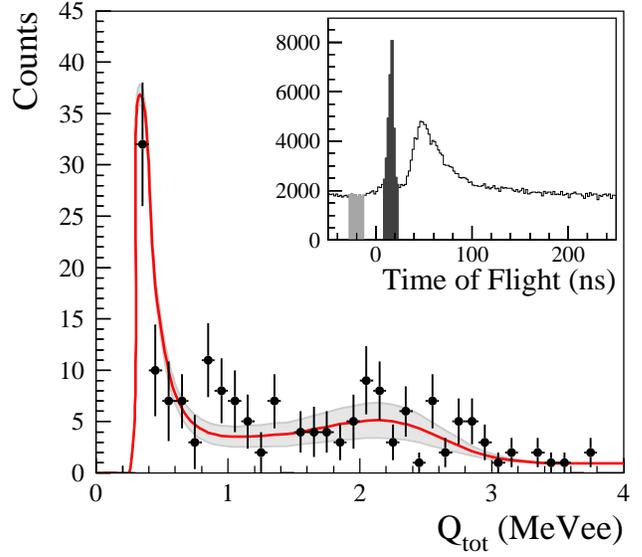}
  \end{center}
  \caption{\small \rm (Color online)  Total charge (Q$_{tot}$) spectrum for prompt $\gamma$-rays 
  recorded in {\sc demon} in coincidence with $^{12}$Be fragments for the C($^{14}$B,$^{12}$Be) reaction.  The line and shaded band represent the best fit (5$\pm$2\%) for the simulated lineshape for 2.1~MeV $\gamma$-rays emitted from the in-flight decay of $^{12}$Be(2$^+_1$).  The inset shows the time-of-flight for events identified as $\gamma$-rays (see text).  Prompt reaction $\gamma$-rays are identified by 
  the dark shaded peak.  The contribution from random coincidences with ambient $\gamma$-rays was estimated as shown by the
  grey shaded region.}
   \label{fig:Gamma}
\end{figure}

Specifically, events corresponding to $\gamma$-rays were selected using the pulse-shape discrimination 
capability of {\sc demon}.  Prompt reaction $\gamma$-rays were then isolated by their time of flight and 
an estimate made of the small number of underlying random events (Inset of Figure~\ref{fig:Gamma}, where for
clarity the time of flight is shown for C($^{14}$B,fragment+$\gamma$)).  
The energy, or total charge(Q$_{tot}$) deposited in the detector modules was carefully calibrated 
using $\gamma$-ray sources and, as such, the form of the Compton edge and overall response to $\gamma$-rays, including
the resolution, was well established \cite{Ran11}.  In order to provide for a comparison
with the data, a simulation was employed to take into account the Doppler 
shift as well as the intrinsic and geometric efficiency of the array for the moving 
source ($\varepsilon_{\gamma}$=0.07\%) \cite{Ran11}.  
Owing to the very limited efficiency for $^{12}$Be+$n$+$\gamma$ triple coincidences ($\sim$0.007\%),
only $^{12}$Be+$\gamma$ coincidences could be examined here.  The results are displayed in Figure~\ref{fig:Gamma}, whereby
the simulated lineshape expected for the decay of $^{12}$Be($2^+_1$) is shown.
The same analysis was also performed for the C($^{14}$B,$^{6}$He+$\gamma$) reaction \cite{Ran11}, whereby, owing to
the lack of any bound $^{6}$He excited states, no events are
expected from a beam velocity source.  This allowed the 
small background of events in the region of Q$_{tot}$ expected for $^{12}$Be$^*$
$\gamma$-decays to be determined and the shape of the fast rising peak of events at threshold to be verified.  
The best fit estimate of the percentage of excited $^{12}$Be($2^+_1$) fragments was
5$\pm$2\% with no discernable population of $^{12}$Be($1^-$).  This result is in line with a measurement
made using a set of small NaI detectors with low efficiency in our
preliminary study of C($^{14}$B,$^{12}$Be+$n$) \cite{Lec02}.  As such, excluding the possible 
population of the isomeric 0$^+_2$ level (Section~V), to which the above analysis is insensitive, the following  discussion will consider that the C($^{14}$B,$^{12}$Be+$n$) reaction proceeds via the $^{12}$Be ground state.

Returning to the interpretation of the C($^{14}$B,$^{12}$Be+$n$) decay energy spectrum, we 
proceed by assuming, as outlined in the Introduction, that the removal of a proton leaves the projectile neutron configuration intact \cite{Zin95,Che01,Lec09,AlF11}.  Given that the 
$^{14}$B valence neutron configuration is dominated by $\nu 2s_{1/2}$ and $\nu 1d_{5/2}$ components \cite{Sau00,Sau04,Gui00,Bed13}
--- $C^2S$($^{13}$B$_{gs}$(3/2$^-$)$\otimes\nu{\it n \ell j}$) = 0.65 and 0.31, respectively \cite{Sau00,Sau04,Gui00} --- 
states with the corresponding neutron configurations are expected to be populated in $^{13}$Be.

\begin{figure}[t!]
 \begin{center}
  \includegraphics[width=\columnwidth]{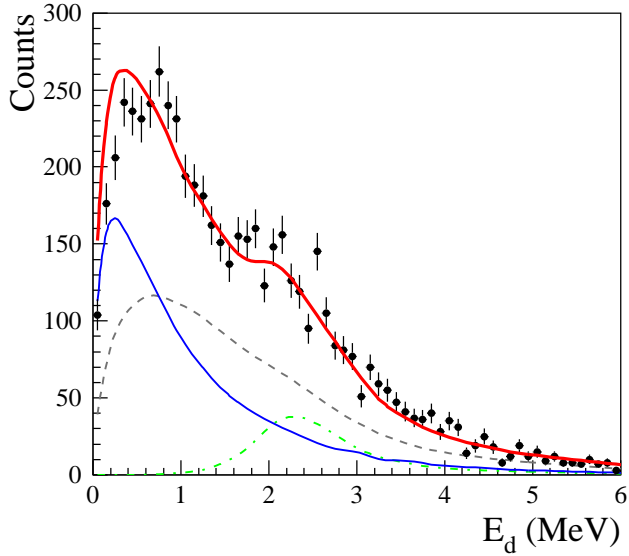}
  \end{center}
  \caption{\small \rm (Color online) Reconstructed $^{12}$Be+$n$ decay energy for the 
  C($^{14}$B,$^{12}$Be+$n$) reaction
  compared to simulations incorporating an $s$-wave virtual state (blue), a higher lying 
  $d$-wave resonance (dot-dashed) 
  and a non-resonant continuum (dashed line) (see text).}
   \label{fig:13Bevirtual}
\end{figure}

\begin{figure}[t!]
 \begin{center}
  \includegraphics[width=\columnwidth]{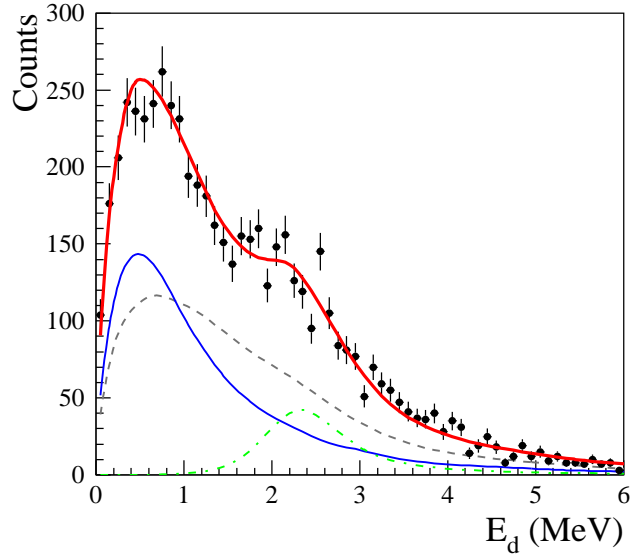}
  \end{center}
  \caption{\small \rm (Color online) Reconstructed $^{12}$Be+$n$ decay energy for the C($^{14}$B,$^{12}$Be+$n$) reaction
  compared to simulations incorporating an $s$-wave resonance (blue), a higher lying 
  $d$-wave resonance (dot-dashed) 
  and a non-resonant continuum (dashed line) (see text).}
   \label{fig:13Besres}
\end{figure}

In order to generate the lineshapes needed to interpret the $^{12}$Be+$n$ decay energy spectrum, the 
relatively simple approach developed in Refs \cite{Zin95,Ber98,Che01} has, as in our earlier 
work \cite{Lec04,Lec09,AlF11}, been followed. 
That is, the lineshape of the fragment--neutron relative
energy distribution was derived from the overlap of the initial bound-state wave function, describing the relative motion between the valence neutron, $\nu{\it n \ell j}$, and the $^{13}$B core
of the $^{14}$B projectile, and the unbound final-state wave function describing the 
$^{12}$Be--neutron interaction (thus neglecting the momentum transfer).  
As detailed in Refs \cite{Lec02,Tho99}, for a resonance,  
this approach results in a lineshape essentially identical to that of
the single-level R-matrix approximation\footnote{As such, possible interference between resonances of the same spin-parity is not considered here.}.  As such, 
R-matrix line shapes were employed \cite{McV94}, whereby the width is dependent on the
energy and $\ell$ through the penetrability 
coefficients, and $\Gamma(E_{d}) = \Gamma_{r} \, P_{\ell}(E_{d})/P_{\ell}(E_{r})$,
where $\Gamma_{r}$ is the width at $E_{d}=E_{r}$ \cite{Gre55}.

In the case of an $s$-wave virtual state, the asymptotic behaviour 
of the bound $\ell=0$ initial-state wave function was described by Hankel functions, with
decay length \linebreak $\alpha=\sqrt{2\mu_{13-n}S_{n}}/\hbar$
(where $\mu_{13-n}$ is the $^{13}$B-$n$ reduced mass and S$_n$ is the $^{14}$B 
single-neutron binding energy of 0.97~MeV).
The asymptotic form of the unbound final state was taken as a linear
combination of spherical Bessel and Neumann functions, characterised 
by the $^{12}$Be-$n$ relative wave number $k=\sqrt{2\mu_{12-n} E_{d}}/\hbar$
(where $\mu_{12-n}$ is the $^{12}$Be-$n$ reduced mass) \cite{Lec02,Ran11}.
A general solution for the unbound scattering state was obtained
by matching the logarithmic derivative of the
asymptotic solution (valid in the region external to the potential) 
and a numerical solution obtained for the internal region \cite{RELISH}.
The scattering length ($a_{s}$), which describes the strength of the fragment-neutron 
final-state interaction in the $s$-wave channel, is related to the phase shift ($\delta_0$) as,
$a_{s}=- \lim_{k \rightarrow 0} \frac{d\delta_{0}}{dk} $. 
From a practical point of view it is important to note that, except for values of $a_{s}$ 
close to zero, the lineshape characteristically rises very quickly at very low E$_d$, then
falls off with a gradual decay towards higher energies \cite{Lec02,Tho99}.

Assuming that the relative strengths of the $s$ and $d$-wave states in $^{13}$Be reflect those
of the parent configurations in $^{14}$B, and given that any $s$-wave state well away from threshold 
will be
extremely broad and consequently difficult to distinguish, the $^{12}$Be+$n$ decay energy spectrum was described in a first instance by a near threshold
virtual $s$-wave state, a higher lying $d$-wave resonance and the uncorrelated distribution.
Apart from the latter (see above), the theoretical lineshapes were used as input to the
Monte Carlo simulation of the experiments.
The scattering length, resonance energy and width, and intensities of each of the three distributions
were left as free parameters.  The best fit to the data is displayed in Figure~\ref{fig:13Bevirtual}, whereby a lower limit on the scattering length of $a_s$=$-$3~fm could be determined and the resonance parameters of the the higher lying feature were deduced to be, 
E$_r$=2.40$\pm$0.20~MeV and $\Gamma_r$=0.90$\pm$0.22~MeV.  Whilst the 
spectrum is well described above 1.5~MeV, the theoretical distribution in the lower region
rises too quickly at threshold and peaks 0.3~MeV below the data.  As in our preliminary work \cite{Lec04},
diminishing, or even eliminating completely the contribution from the uncorrelated events degrades further the agreement with the data.  It may also be noted that an even worse description of the data is obtained if
the very strongly interacting virtual $s$-state ($a_{s}$$<$$-$10~fm) posited in the stable beam fragmentation experiments  is employed \cite{Tho00,Chr08}.  We shall return to this in the following section.

\begin{figure}[t!]
 \begin{center}
  \includegraphics[width=\columnwidth]{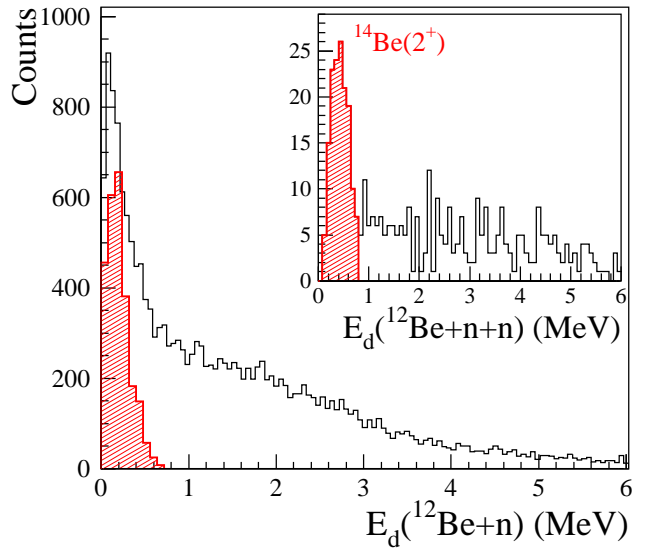}
  \end{center}
  \caption{\small \rm (Color online) Main panel: Reconstructed $^{12}$Be+$n$ decay energy for the C($^{15}$B,$^{12}$Be+$n$) reaction.  Inset: the 
  $^{12}$Be+$n$+$n$ decay-energy spectrum from the C($^{15}$B,$^{12}$Be+$n$+$n$) reaction.  The red cross-hatched distribution in the main panel
  represents $^{12}$Be+$n$ events, reconstructed from the $^{12}$Be+$n$+$n$ data, arising from the decay of the $^{14}$Be(2$^+_1$) state (cross hatched events in inset) and renormalised to account for the one versus two-neutron detection efficiencies. }.
   \label{fig:15B12Ben}
\end{figure}

At this juncture it is worth recalling that the conditions required for the formation
of a virtual $s$-wave scattering state are quite specific --- namely, the potential 
should be spherical and inert with no internal degrees of freedom (as such the scattering 
state will be by definition pure $s$-wave).  Given that $^{12}$Be is known to be 
well-deformed \cite{Iwa00}, the formation of resonances with $\ell_n$=0 becomes possible \cite{Joa65,Boh69}.
Indeed, within the R-matrix approach resonances and virtual $s$-states occur depending on the character of the corresponding poles \cite{McV68,McV94}.  Thus, in the same vein as our preliminary study \cite{Lec04}, a description of the
$^{12}$Be+$n$ decay energy spectrum employing an $s$-wave resonance at low energy was attempted.  
Figure~\ref{fig:13Besres}
displays the results, whereby a good agreement is obtained with the data for an $s$-wave resonance at 0.70$\pm$0.11~MeV and $\Gamma_r$=1.70$\pm$0.22~MeV, along with a $d$-wave resonance at 2.40$\pm$0.14~MeV with $\Gamma_r$=0.70$\pm$0.32~MeV.  The uncertainties include, not only those arising from the fitting, but also an estimate of effects arising from the calibrations
(in particular the absolute calibrations of the $^{12}$Be and neutron momenta) and the construction of the uncorrelated distribution.  In the context of the latter, it may be noted that 
the description of the data in the region below $\sim$1~MeV is rather poor if the uncorrelated distribution is eliminated \cite{Ran11}.

\subsection{The C($^{15}$B,$^{12}$Be+$n$) reaction}  

The reconstructed decay energy spectrum for the C($^{15}$B,$^{12}$Be+$n$) reaction is displayed in
Figure~\ref{fig:15B12Ben}.  The most striking aspect of the spectrum, which bears little resemblance to that obtained in
proton removal from $^{14}$B, is the very narrow, strongly populated peak at
threshold.  The removal of a proton and a neutron from
$^{15}$B may occur in two ways: {\it (i)} both particles are directly removed in a single step by the interaction with the target, or {\it (ii)} in a first step, proton removal populates
continuum states in $^{14}$Be which subsequently neutron decay to $^{13}$Be\footnote{Owing to the very
much higher (16~MeV) proton versus neutron separation energy in $^{14}$B, neutron removal followed by
proton decay is extremely unlikely.}.  In the first case, the states in $^{13}$Be are directly populated, whilst the character of any intermediate states in the second scenario will influence to some degree
the $^{12}$Be+$n$ decay energy spectrum.  Such effects were evaluated here by analysing the 
$^{12}$Be+$n$+$n$ events. 

\begin{figure}[t!]
 \begin{center}
  \includegraphics[width=\columnwidth]{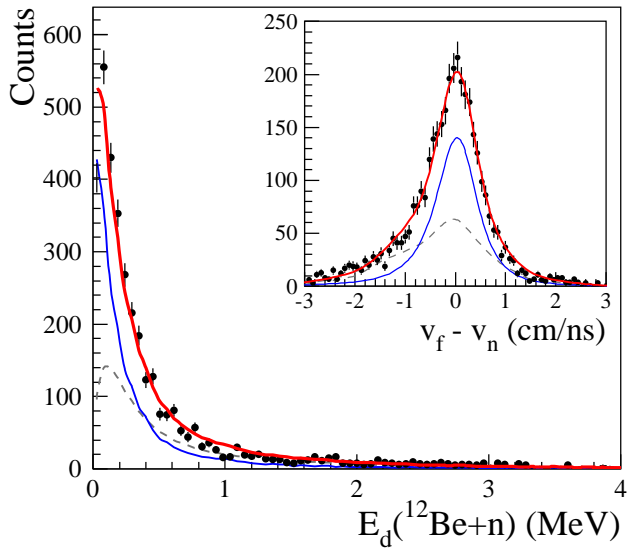}
  \end{center}
  \caption{\small \rm (Color online) Reconstructed $^{12}$Be+$n$ decay energy and relative velocity (inset) spectra from
  the C($^{15}$B,$^{12}$Be+$n$) reaction for neutrons with $\theta_{lab}<$ 3$^{\circ}$.  Both the uncorrelated
  distribution (dashed) and that representing an $s$-wave virtual state with $a_s$=$-$10~fm (blue line) 
  are shown.}
   \label{fig:15B12BenEdvrel3}
\end{figure}

The inset of Figure~\ref{fig:15B12Ben} shows the reconstructed $^{12}$Be+$n$+$n$ decay energy spectrum.  As expected the statistics are reduced with respect to the $^{12}$Be+$n$ data owing to the smaller two-neutron detection efficiency, which includes the cross-talk rejection (Section~II).  Nevertheless the spectrum displays, in addition to a very broad non-resonant continuum, a prominent peak some 0.3~MeV above the two-neutron threshold which corresponds to the known E$_x$=1.5~MeV 2$^+_1$ state of $^{14}$Be \cite{Sug07}.
Clearly, given the limited energy available for the decay of this state into $^{12}$Be+$n$+$n$, 
$^{12}$Be+$n$ events of even lower average decay energy will be observed.  
Indeed, when  
the $^{12}$Be+n events are confined to those arising from the decay of $^{14}$Be(2$^+$) --- E$_{12Be+n+n}<$0.8~MeV --- they are seen, unsurprisingly, to constitute the very narrow structure at 
threshold (Figure~\ref{fig:15B12Ben}).  That is to say, the threshold strength in the $^{12}$Be+$n$ relative energy spectrum arises 
from the limited energy available in the decay of $^{14}$Be$^*$(2$^+$) populated in the proton removal from $^{15}$B. 
Given that the form and threshold location of the peak 
is very close to that associated with a 
strongly interacting $s$-wave
virtual state, it is very probable that the $^{12}$Be+$n$ coincidence 
data from stable beam fragmentation was dominated by the decay of $^{14}$Be(2$^+$) rather than 
direct population of $^{13}$Be, as was implicitly assumed in the analyses of these experiments \cite{Tho00,Chr08}. The present work mirrors that of the earlier study of 
Kondo {\it et al.} \cite{Kon10}, whereby the population of $^{14}$Be(2$^+$) in the dissociation of $^{14}$Be was identified, also through the analysis of the $^{12}$Be+$n$+$n$ events, as the source of the sharp threshold peak in the corresponding $^{12}$Be+$n$ decay energy spectrum.

It is possible to proceed further and mockup the very restricted acceptances of the stable beam studies by
limiting the neutron detection in the present work to very forward angles ($\theta_n<$3$^\circ$) and
analysing the resulting $^{12}$Be+n events.  
In Figure~\ref{fig:15B12BenEdvrel3} the
results are shown for the $^{12}$Be+$n$ decay energy and relative velocity, whereby the 
restricted acceptances privilege almost exclusively the peak at threshold produced by the 
decay of $^{14}$Be$^*$(2$^+$) and resemble very closely the
Be($^{18}$O,$^{12}$Be--$n$) relative velocity spectrum of Ref.~\cite{Tho00} and the
Be($^{48}$Ca,$^{12}$Be--$n$ decay energy spectrum of Ref.~\cite{Chr08}. 

To provide a more quantitative comparison, the restricted acceptance data set was used to generate uncorrelated $^{12}$Be--$n$ distributions and the two spectra were then described using 
this distribution (which, as noted earlier, includes the experimental response) together with a virtual $s$-wave state lineshape that had been filtered through the simulation \cite{Lep09}. 
In order to make a direct comparison, the results are shown for the limit on the scattering length of 
--10~fm deduced in Ref.~\cite{Tho00}.   As may be seen, such a description 
reproduces very well both spectra and reinforces the notion that the strongly interacting 
$s$-wave virtual state deduced from the stable beam
fragmentation studies was an artifact resulting from the population and decay of the $^{14}$Be(2$^+$) state.
In the following, therefore, the discussion concentrates on the interpretation, in the light of theoretical
considerations, of the C($^{14}$B,$^{12}$Be+$n$) decay-energy spectrum. 

\section{Discussion}

As noted earlier, single-proton removal from $^{14}$B is expected to populate final states in $^{13}$Be
with neutron configurations corresponding to those of the valence neutron in the former --- that is, 
$\nu 2s_{1/2}$ and $\nu 1d_{5/2}$ configurations.  In this context (Section  III-A), the broad
structure at 0.70~MeV may be identified as a 1/2$^+$ state and that at 2.40~MeV as 5/2$^+$. In terms of the latter,
the single-particle width for a $d$-wave resonance at this energy is $\Gamma_{sp}\approx$0.90~MeV to be compared to
a measured width of 0.70$\pm$0.32~MeV, suggesting a strong single-particle character.  In a simple picture, whereby the 
$^{13}$Be levels are essentially single-particle ($^{12}$Be $\otimes$ $\nu nlj$), yields reflecting the spectroscopic
factors characterising the initial state would be expected for proton removal \cite{Lec09}.  After correcting for the detection efficiency and the contribution of the uncorrelated distribution, the structure centred around 0.70~MeV is seen to be populated 3.6$\pm$0.8 times more strongly than that at 2.40~MeV,
which may be compared to the ratio of 2.1 of the corresponding $^{14}$B spectroscopic 
factors (Section  III-A).  Finally, the separation between the two levels of 1.70$\pm$0.18~MeV may be compared to the
1/2$^+$--5/2$^+$ energy difference of some 2.3~MeV estimated \cite{For13} on the basis of a linear extrapolation 
of the corresponding levels in less exotic N=9 isotones, in the spirit of 
Talmi and Unna \cite{Tal60}, and the difference in the $\nu 2s_{1/2}$ and $\nu 1d_{5/2}$ effective single-particle energies of around 1.5~MeV in $^{14}$B deduced from a very recent d($^{13}$B,p) reaction study \cite{Bed13}.
Interestingly, it may be noted that the lowering of the $\nu 2s_{1/2}$ orbital with respect to the $\nu 1d_{5/2}$
is expected in a simple potential \cite{Mil01} and is enhanced by weak binding \cite{Tan96}, effects which have been revisited in a very recent study of the systematics of the 1/2$^+$ and 5/2$^+$ levels in light nuclei \cite{Hof13}.

\begin{figure}[t!]
 \begin{center}
  \includegraphics[width=\columnwidth]{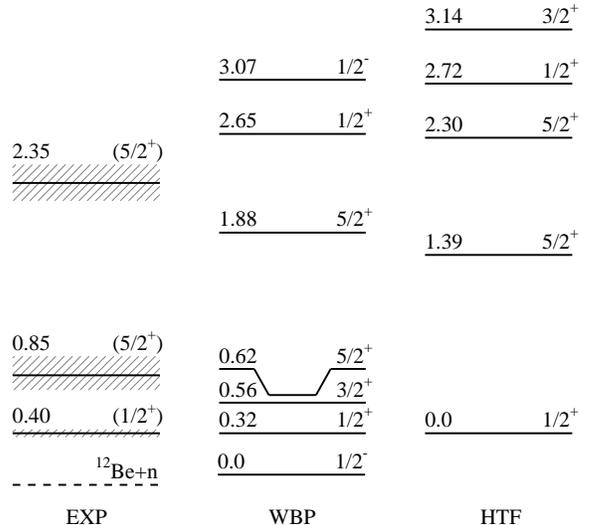}
  \end{center}
  \caption{\small \rm Levels (MeV) in $^{13}$Be as predicted by (0-3)$\hbar\omega$ shell 
  model calculations (WBP) and estimated, for positive parity states, within a simplified scheme by Fortune (HTF) \cite{For13} where the lowest 1/2$^+$ state
  is assumed to lie 0.4~MeV above threshold.  These two level schemes are normalised to the
  predictions for the 1/2$^+_1$ state and the energies indicated are with respect to the lowest-lying level of each calculation.
  The results of the present work are also shown (EXP), where the level 0.40~MeV above the $^{12}$Be+$n$ threshold is
  identified with the predicted 1/2$^+_1$ state.  The experimental energies are listed with respect to the
  $^{12}$Be+$n$ threshold.  The shaded bands represent the experimental uncertainties in the resonance energies (Table~II).}
   \label{fig:levelscheme}
\end{figure}

Given the mixed (0+2)$\hbar\omega$ character of $^{12}$Be \cite{Bar76,Nav00,Pai06}, it is clear that interpretations beyond the naive $^{12}$Be$\otimes\nu nlj$ description should be sought.  In this context, Fortune \cite{For13} has estimated, within a simple model \cite{For10}, the energies of the 
$^{10}$Be$\otimes$($\nu 2s1d$)$^3_J$ levels with particular focus on the corresponding 5/2$^+$ state.  The results are shown in Figure~\ref{fig:levelscheme}, where it is assumed that the
lowest 1/2$^+$ level is 0.4~MeV above threshold, as will be proposed here (see below).  The lowest-lying 5/2$^+$ level, some 1.4~MeV above 
the 1/2$^+$ $^{12}$Be$\otimes \nu 2s_{1/2}$
state, is $^{10}$Be$\otimes(\nu 2s1d)^3$, whereas the 5/2$^+$ $^{12}$Be$\otimes\nu 1d_{5/2}$ level is, based on the estimated 1/2$^+$--5/2$^+$ energy difference, higher lying at 2.3~MeV.  In terms of the interpretation of the present data, unless the 1/2$^+$ state were to lie
at threshold, the 5/2$^+_2$ level is energetically permitted to decay via $d$-wave neutron emission to the isomeric $^{12}$Be(0$^+_2$) level (E$_x$=2.24~MeV), the decay of which could not be detected here.  
As noted by Fortune and Sherr \cite{For10}, despite the small available decay energy, 
structural considerations suggest that decay to the isomeric state could possibly be favoured (we shall return to this point below).

In order to explore in more detail the location and structure of the low-lying states of $^{13}$Be, shell model
calculations within the $s$-$p$-$sd$-$pf$ model space and employing the WBP interaction \cite{WBP} have been undertaken.
The calculations follow the same prescription as that of Ref. \cite{Kan10}, whereby an additional $psd$ energy-gap
parameter ($\Delta_{spd}$) was introduced to account for the mixing between the 0 and 2$\hbar\omega$ and 1 and 3$\hbar\omega$
configurations \cite{War92}.  Importantly, $\Delta_{spd}$ was chosen in order to 
reproduce the experimentally determined 
configuration mixing in the $^{12}$Be$_{gs}$ \cite{Nav00,Pai06}.  The resulting level scheme (up to 3~MeV) for $^{13}$Be is displayed in Figure~\ref{fig:levelscheme}, whilst the energies, spectroscopic factors for proton removal from $^{14}$B and those describing the structural overlap with states in $^{12}$Be are tabulated in
Table~I. 

Whilst our interest here is primarily on the positive parity states, it is interesting to note that the 
1/2$^-$ 1$\hbar\omega$ $\nu 1p_{3/2}^{-1}$  level is predicted to be the lowest lying one, 0.32~MeV below
the 1/2$^+_1$ state.  The first 5/2$^+$ level with a 2$\hbar\omega$ $^{10}$Be$\otimes$($\nu 2s1d$)$^3$ 
parentage, is found at 0.62~MeV, in close proximity to the 3/2$^+$ level with the same parentage.
The second 5/2$^+$ level with a 0$\hbar\omega$ $^{12}$Be$\otimes$($\nu 1d_{5/2}$) parentage
is predicted to lie 1.88~MeV above the 1/2$^-$, implying a 1/2$^+$--5/2$^+$ energy difference of 1.56~MeV --- 
smaller than that based on the extrapolation from less exotic N=9 isotones.  Interestingly, in unmixed 
calculations the 5/2$^+$ 0 and 2$\hbar\omega$ levels are degenerate and the 0$\hbar\omega$ 1/2$^+$ state 
lies only around 0.7~MeV below and is separated by 1.7~MeV from its higher lying 2$\hbar\omega$ counterpart \cite{Mil13}.   

\begin{figure}[t!]
 \begin{center}
  \includegraphics[width=\columnwidth]{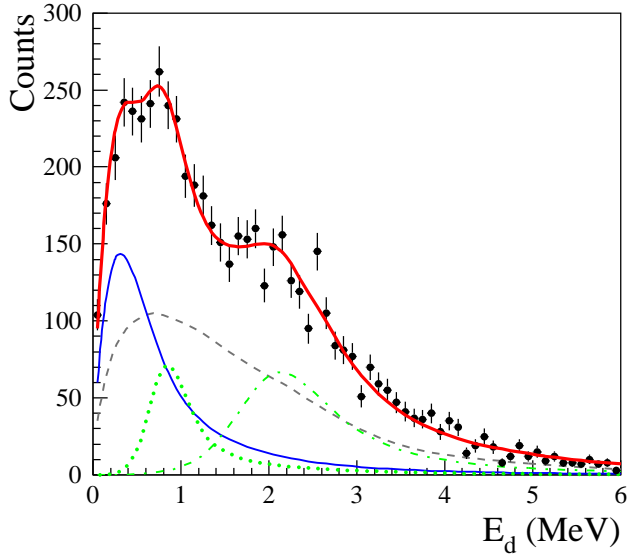}
  \end{center}
  \caption{\small \rm (Color online) Reconstructed $^{12}$Be+$n$ decay energy for the C($^{14}$B,$^{12}$Be+$n$) reaction
  compared to simulations incorporating low-lying $s$ (blue) and $d$-wave resonances (dotted), and a
  higher lying 
  $d$-wave resonance (dot-dashed) together with a non-resonant continuum (dashed line).}
   \label{fig:13Besdd}
\end{figure}

The most strongly populated states in single-proton removal from $^{14}$B are 
predicted (Table~I) to be the 1/2$^+_1$ ($C^2S$=0.41) and
5/2$^+_2$ states ($C^2S$=0.43), whilst the 5/2$^+_1$ level is 
predicted to carry a factor 3 less strength ($C^2S$=0.13).
The remaining strength --- less than 5\% of the total lying below the two-neutron 
emission threshold --- is carried by the
1/2$^+_2$ level ($C^2S$=0.05).  As the 3/2$^+$ 2$\hbar\omega$ state does not have a 
corresponding 0$\hbar\omega$ level to mix with, it will be only very weakly populated in 
direct proton removal from $^{14}$B.
In the case of the 1/2$^+$ states, the quite large separation permits only a relatively small amount
of mixing to occur and hence a rather weak yield to the 1/2$^+_2$ level.

In the light of these calculations, the description of the C($^{14}$B,$^{12}$Be+$n$) decay 
energy spectrum has been reappraised
under the assumption of three states being appreciably populated --- low lying (E$_d$$\lesssim$1~MeV) 
$s$ and $d$-wave levels together with a higher lying (E$_d$$\sim$1.5--3.0~MeV) $d$-wave
resonance.  As may be seen in Figure~\ref{fig:13Besdd}, the spectrum is very well reproduced, including,
in particular, the region below 1~MeV, which is 
composed of an $s$-wave resonance at 0.40$\pm$0.03~MeV ($\Gamma_r$=0.80$^{+0.18}_{-0.12}$~MeV) and a narrower
$d$-wave resonance at 0.85$^{+0.15}_{-0.11}$~MeV ($\Gamma_r$=0.30$^{+0.34}_{-0.15}$~MeV), whilst the higher lying $d$-wave resonance is at E$_r$=2.35$\pm$0.14~MeV with a width $\Gamma_r$=1.50$\pm$0.40~MeV (Table~II and Figure~\ref{fig:levelscheme}).  The widths of the $d$-wave states may be compared to estimates of the corresponding single-particle widths of 0.08 and 0.90~MeV.
We note that the experimental 
resolution ({\sc fwhm}) at 0.85~MeV is $\sim$380~keV, hence the uncertainty on the width of this level
is rather large.

In terms of the yields to the three levels, the shell model calculations indicate, normalising the 1/2$^+$ to a strength of 1.00,
that the 5/2$^+_1$ and 5/2$^+_2$ states should be populated with strengths of 0.32 and 1.05.  Experimentally, 
strengths of 0.40$\pm$0.07 (5/2$^+_1$) and 0.80$\pm$0.09 (5/2$^+_2$) are observed with respect to the 1/2$^+$ 
level (Table~II), in reasonably good agreement with the shell-model calculations.  

Energetically, only the level at 2.35~MeV may decay to excited states in $^{12}$Be (we ignore the possible small yield to the 1/2$^+_2$ level) rather than only directly to the ground state.  In the case of decay
to the isomeric 2.25~MeV $^{12}$Be(0$^+_2)$ state, this can only occur for the high-energy side of the
2.35~MeV resonance and, despite the large structural overlap (Table~I), will be very strongly suppressed by the
very low effective decay energy.  Decay to the 2.11~MeV $^{12}$Be(2$^+_1)$ state is slightly more favourable as energetically a larger fraction of the resonance can decay via this pathway.  Structurally there is
a significant overlap for decay via $s$-wave neutron emission to $^{12}$Be(2$^+_1)$ and, whilst the
decay to the $^{12}$Be$_{gs}$ is energetically greatly favoured, it is structurally very strongly suppressed (Table~I).
Experimentally, however, no indication was found here (Section~III-A) for a significant yield to 
$^{12}$Be(2$^+_1)$.  In addition, from a practical point of view any low energy decays to $^{12}$Be(2$^+_1)$
(or $^{12}$Be(0$^+_2)$) would need to be quite strong and narrow to be identified at 
threshold in the decay energy spectrum.  We note that if the energy of the proposed 5/2$^+_2$ level is at 
the lower bounds of the present uncertainties, then the fraction of the distribution energetically allowed to 
decay via neutron emission to the $^{12}$Be(2$^+_1)$ state would be greatly reduced.  Alternatively, or 
in addition, it is
possible that the structural overlap between the 5/2$^+_2$ level and the $^{12}$Be(2$^+_1)$ state is
overestimated by the shell model.

\begin{table}[t!]
   \begin{center}
       \begin{tabular}{cccccc}
        \hline \hline
         $J^{\pi}_i$ & $E_x$ (MeV) & $C^2S$ & $b_{s/d}$  & $b_{s/d}$  & $b_{s/d}$  \\ 
                     &             &        &  $^{12}$Be(0$^+_1)$ &  $^{12}$Be(2$^+_1)$ &  $^{12}$Be(0$^+_2)$ \\ \hline
          1/2$^-_1$   & 0.0   &       &   &   &   \\
          1/2$^+_1$   & 0.316 & 0.41  & 0.57($s$)  & 0.05($d$)  & 0.23($s$)  \\
          3/2$^+_1$   & 0.562 & 0.00  & 0.04($d$)  & 1.13($d$)  & 0.01($d$)  \\
          5/2$^+_1$   & 0.619 & 0.13  & 0.67($d$)  & 0.08($s$)/0.05($d$)  &  $<$0.01 \\
          5/2$^+_2$   & 1.885 & 0.43  & 0.01($d$)  & 0.23($s$)/0.01($d$)  &  0.65($d$)  \\
          1/2$^+_2$   & 2.652 & 0.05  & 0.05($s$)  & 0.33($d$)  &  0.35($s$) \\
          1/2$^-_2$   & 3.069 &       &   &   &   \\  \hline
       \hline
       \end{tabular}
       \caption{\small \rm Low-lying level structure of $^{13}$Be predicted 
       by (0-3)$\hbar\omega$ shell-model calculations using the WBP interaction \cite{War92} in the $s$-$p$-$sd$-$fp$ valence space (see text).
         $E_x$ is the excitation energy with respect to the 1/2$^-_1$
        state; $C^2S$ is the spectroscopic factor for removing a 1$p_{3/2}$ proton from $^{14}$B; 
        $b_{s/d}$ is the spectroscopic factor for $s$ or $d$-wave neutron decay to $^{12}$Be(0$^+_1$,2$^+_1$,0$^+_2$)
        for states with non-zero $C^2S$ (values are listed even if the decay is not energetically possible). }
      \label{tab:sm}
   \end{center} 
\end{table}

\begin{figure}[t!]
 \begin{center}
  \includegraphics[width=\columnwidth]{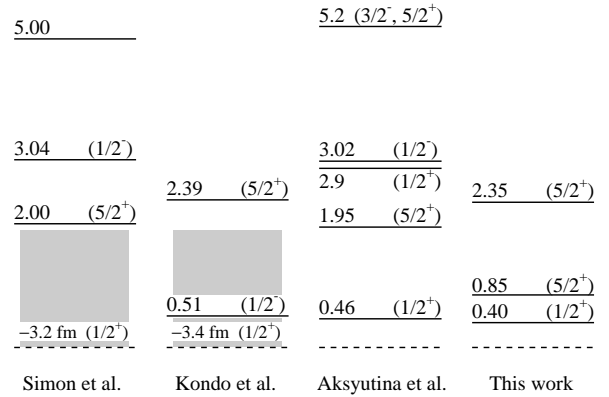}
  \end{center}
  \caption{\small \rm (Color online) Comparison of the present work with the neutron removal studies of
  Simon \etal ~\cite{Sim07}, Kondo \etal ~\cite{Kon10} and Aksyutina \etal ~\cite{Aks13a}.  The energies of
  the levels (MeV) with respect to the $^{12}$Be+$n$ threshold (dashed line) are shown, together 
  with the proposed spin-parity assignments.  In the case of Refs \cite{Sim07,Kon10} weakly interacting
  virtual $s$-wave strength is indicated by the grey bands and the corresponding scattering lengths 
  are noted.}
   \label{fig:13Becomp}
\end{figure}

In closing the Discussion, we turn to a comparison with recent work 
which has employed the complementary probe of neutron removal from $^{14}$Be \cite{Sim07,Kon10,Aks13a}.
The results of these studies have been enumerated in Section~I and are displayed in
Figure~\ref{fig:13Becomp} together with the present work.
As already observed in Ref.~\cite{Aks13a}, whilst the decay-energy spectra are similar --- a 
well populated
peak-like structure centred below 1~MeV, with a less well defined feature
at higher energy ($\sim$2--3~MeV) --- the interpretations
are at variance.     

Both the present study and the latest neutron knockout work \cite{Aks13a} agree that  
the low-energy region of the decay energy spectra 
are dominated by resonant $s$-wave strength rather than 
the weakly interacting virtual
$s$-wave strength invoked earlier \cite{Sim07,Kon10}.  Whilst Ref.~\cite{Aks13a} employed the contention 
of our earlier $^{14}$B proton-removal study of a broad low lying $s$-wave 
resonance \cite{Lec04} (explored here in Section~III-A and Figure~\ref{fig:13Besres}) as initial input to their analysis, the dominance of the $s$-wave strength was confirmed by the $^{12}$Be+$n$
``profile'' function (transverse momentum distribution as a function of decay energy) \cite{Aks13,Aks13a}.
In contrast, the intermediate energy measurement \cite{Kon10} identified the low-lying peak as  
an $\ell$=1 resonance based on the associated $^{12}$Be+$n$
transverse momentum distribution. 
The strength  --- almost 50\% of the total measured yield --- is somewhat surprising given that the $^{14}$Be wavefunction is believed to be dominated by $\nu 2s_{1/2}^2$ 
and $\nu 1d_{5/2}^2$ valence neutron configurations \cite{Suz99,Lab01}.  It may be noted that the
form of the decay energy spectrum near threshold in this measurement is dependent 
on the subtraction of the narrow component arising from the population and decay 
of the $^{14}$Be($2^+_1$) (Section~III-B).

In terms of $s$-wave strength, it was also proposed \cite{Aks13a}, following
structural considerations similar to those discussed above of Fortune \cite{For13} (Figure~\ref{fig:levelscheme}), that a second higher lying 1/2$^+$ (2.9~MeV) is populated in
the high-energy neutron removal.  Whilst difficult to identify directly in the decay energy spectrum owing to
its very broad width ($\Gamma$$\approx$4~MeV), its 
presence was argued for on the basis of the influence on the lineshape of the combined $s$-wave strength.  
Here, as noted earlier, in proton removal from $^{14}$B,
the 1/2$^+_2$ --- predicted by the shell model at around 2.7~MeV\footnote{Assuming the
1/2$^+_1$ level to lie at 0.40MeV (Figure~\ref{fig:levelscheme}).} --- is expected to be only relatively 
weakly populated (Table~I), whereas the identification of
such a state in the present data would require a significant yield to it
given the intrinsically very broad width and the proximity of the 2.35~MeV level.

Turning to the $d$-wave strength, all of the studies agree on its presence in the region of $\sim$2--2.5~MeV.  
Whilst the present investigation and that of Kondo \etal ~\cite{Kon10} concur on the energy of the
corresponding 5/2$^+$ level, the high energy neutron removal studies, guided by 
the heavy-ion multi-nucleon transfer and pion absorption 
experiments \cite{Ale83,Ost92,Bel98,Gor98}, place it around
0.4~MeV lower \cite{Sim07,Aks13a}.   
In contrast, none of the neutron removal studies observe any clear evidence for a low-lying 5/2$^+$
level, such as that deduced in the present work.  Ref.\cite{Aks13a}, however, suggested,
based on the profile function, that there is the possibility that an $\ell >$0 resonance might
exist at around 1~MeV.  Conceivably, depending on the detailed structure of $^{14}$Be, such
a state might be populated with much less strength than in the case of proton removal from $^{14}$B.

Above 2.5~MeV, the present study and that of Ref.~\cite{Kon10} see no clear
evidence for any states (note that for levels above 3.17 and 3.67~MeV the $^{11}$Be+2$n$ and 
$^{10}$Be+3$n$ channels are open).  The high-energy neutron knockout studies \cite{Sim07,Aks13a},
however, propose levels at 3.0 and close to 5~MeV, as inferred from the multi-nucleon transfer and pion absorption 
experiments \cite{Ost92,Bel98,Gor98}.  The former was assigned 1/2$^-$ and the 
latter suggested to be 3/2$^-$ or 5/2$^+$ \cite{Aks13a}.  Experimentally these 
two levels are very broad and 
quite weakly populated and the presence of at least the 5~MeV level should probably 
be regarded as provisional.

In terms of $p$-wave strength, only the 
intermediate energy neutron removal study claims to observe a low-lying (E$_r$=0.51~MeV)
1/2$^-$ resonance \cite{Kon10}.  As already noted, neither of the high-energy
studies corroborate this result, whilst the present investigation should not populate
$p$-wave strength in any appreciable manner.  As argued in Section~I, there is good evidence from multi-nucleon transfer reaction 
\cite{Bel98} and pion absorption studies \cite{Gor98} for the existence of a 1/2$^-$ level at around 0.8~MeV.  In addition, the shell model calculations predict (Figure~\ref{fig:levelscheme} 
and Table~I) that the 1/2$^-_1$ level should appear at low energy.
As such it may be concluded that the lowest 1/2$^+$ and $1/2^-$ levels lie close together below
1~MeV.

\begin{table}[t!]
   \begin{center}
       \begin{tabular}{ccccccc}
        \hline \hline
    $J^\pi$  \,   & $E_r$ (MeV) \, & \, $\Gamma_r$ (MeV)\,  & \, $I/I(1/2^+)$ \, \\ 
 \hline
 $1/2^+$		& 0.40$\pm$0.03  		&  0.80$^{+0.18}_{-0.12}$ 	& 1.00		\\
 $5/2^+$  	& 0.85$^{+0.15}_{-0.11}$   	&  0.30$^{+0.34}_{-0.15}$	& 0.40$\pm$0.07	\\
 $5/2^+$  	& 2.35$\pm$0.14 		&  1.50$\pm$0.40  		& 0.80$\pm$0.09	\\ 
       \hline \hline
       \end{tabular}
       \caption{\small \rm Deduced $J^\pi$ assignments, resonance energies ($E_r$) above 
       the $^{12}$Be+$n$ threshold, 
       widths ($\Gamma_r$) and strengths with respect to the 1/2$^+$ level,
       for the adjustment to the data shown in Figure~\ref{fig:13Besdd}.}
      \label{tab:exp_sdd}
   \end{center} 
\end{table}

\section{Conclusions}

In summary, the low-lying structure of the neutron unbound system $^{13}$Be has been explored 
via invariant mass spectroscopy of the beam velocity neutron and $^{12}$Be charged fragments from reactions
of intermediate energy (35~MeV/nucleon) secondary beams of $^{14,15}$B on a carbon target.   
In the case of the breakup of $^{15}$B, a very sharp peak at threshold was observed in the 
$^{12}$Be+$n$ decay energy spectrum. 
Analysis of the $^{12}$Be+$n$+$n$ events demonstrated, in line with earlier work \cite{Kon10}, that 
this resulted from the sequential-decay of the unbound $^{14}$Be(2$^+$) state
rather than a strongly interacting $s$-wave virtual state in $^{13}$Be as had been 
surmised in some earlier stable beam fragmentation studies. 

In a second measurement, single-proton removal from $^{14}$B was investigated.  In this case 
a broad low-lying structure some 0.7~MeV above
the neutron-decay threshold was observed in the $^{12}$Be+$n$ decay energy spectrum in addition to a less prominent feature at around 2.4~MeV. 
Based on the selectivity of the reaction, which should populate states in $^{13}$Be with neutron
configurations mirroring those of $^{14}$B, and a comparison with (0-3)$\hbar\omega$ shell-model calculations, the low-lying structure was deduced to most probably arise from J$^\pi$=1/2$^+$ and 5/2$^+$ resonances at 0.40$\pm$0.03 
and 0.85$^{+0.15}_{-0.11}$~MeV
above threshold. The higher lying feature is believed to arise from a second broad 5/2$^+$ level 
at 2.35$\pm$0.14~MeV.
Taken in conjunction with earlier studies and the shell model calculations presented here, the lowest 1/2$^+$ and 1/2$^-$ levels would appear to lie relatively 
close together below 1~MeV.

A number of improvements are clearly possible in terms of the work presented here.  First, a more
granular neutron array would enable the resolution in the reconstructed $^{12}$Be+$n$ decay energy
to be improved, and thus allow the presence of the two low-lying resonances  
to be put on a firmer footing.  Second, the introduction of a dedicated $\gamma$-detection array with
good efficiency would allow any possible decays via the $^{12}$Be(2$^+_1)$ state to be investigated 
in coincidence with the neutrons for even relatively weak decays.  Unfortunately, improved coincident
$\gamma$-ray detection will
not eliminate the complications arising from possible decays via the isomeric $^{12}$Be(0$^+_2)$ state.  
Finally, a realistic theoretical treatment of the reaction would be welcome, including the population
of the non-resonant continuum.  Whilst this is not an easy task, some notable steps in this direction have already been taken (see, for example, Ref.~\cite{Bla07}).

Ideally, neutron transfer onto $^{12}$Be should be studied using the (d,p) reaction (in inverse kinematics)
whereby no ambiguity 
will arise in the energies of the states populated as they are derived from the energies and angles of the protons.
Moreover, such an approach should in principle populate both positive and negative parity
states with $^{12}$Be$_{gs}\otimes\nu nlj$ components. Such an experiment presents 
many challenges, not the least of which is the $^{12}$Be beam, the
energy of which must be relatively low (on the order of 5$-$10~MeV/nucleon) in order to provide for reasonable cross
sections for the low angular momentum transfers of interest.


\begin{acknowledgments}

The support provided by the technical staff of {\sc lpc} and the {\sc lise} crew is gratefully acknowledged, 
as are the efforts of the {\sc ganil} cyclotron operations team for providing the primary beam.  Enlightening exchanges with John Schiffer
and John Millener are also acknowledged.
We also express our appreciation for the vital contributions made by our late colleague and friend Jean-Marc Gautier to all the {\sc charissa+demon} experiments undertaken by our collaboration.  
This work has been supported in part by the European Community within the FP6
contract EURONS RII3--CT-2004-06065.  BAB acknowledges support from NSF grant PHY-1068217.

\end{acknowledgments}

\bibliography{13be_bibliography_new}

\end{document}